# An Inertial Reaction to Cosmological Accelerations

*Scott Funkhouser*
*Occidental College, Los Angeles CA, 90041, USA*
5May2005

**ABSTRACT**
Mach's "fixed stars" are actually not fixed at all. The distant clusters of galaxies are not only receding from each observer but they are also accelerating since the rate of cosmological expansion is not constant. If the distant cosmic masses in someway constitute the frame of inertial reference then an additional force should be generated among local bodies in reaction to the apparent cosmological accelerations of the distant galaxies.

Newton's Second Law is incomplete without a frame against which acceleration is to be defined. According to Mach's principle the bulk of cosmic mass constitutes the frame relative to which acceleration produces an inertial reaction. Mach hypothesized that there was some yet-unspecified acceleration-dependent induction among masses that cumulatively generates inertia when a body is accelerated with respect to the bulk of the mass of the Universe. Even in relativistic cosmological models the large-scale distribution of matter may constitute the frame that determines the evolution of the axis of a gyroscope [1].

Mach's conception of the cosmos was formed before Hubble elucidated the nature of the Andromeda nebula and discovered the redshifts in the galactic spectra. Mach treated the "stars" of the Universe as "fixed", constituting a static reservoir of mass. This presumption must be modified to meet the modern understanding of the cosmos. All of space has been expanding since the Big Bang, so the distant clusters of galaxies are all receding, moving away radially with respect to any point in the Universe. This in itself does not necessarily pose a problem for Mach's Principle since any interaction responsible for inertia should depend on relative accelerations between masses and not relative velocities. However, the expansion of the Universe has not occurred at a constant rate. The distant galaxies and quasars are not just receding but receding at a rate that is changing and are therefore *accelerating* with respect to any point in space.

Not all motion associated with the expanding Universe can be expected to result in a physical interaction since the cosmological recession includes at least one component of motion that is not "real" motion in the mechanical sense. Galaxies with redshifts greater than ~3 are receding at superluminal velocities, but this is not inconsistent with the special theory of relativity [2]. The total velocity of a distant galaxy relative to a local observer consists of two terms: a peculiar velocity that produces a special-relativistic Doppler shift and a recessional velocity that produces a redshift through the expansion of space. The peculiar velocity $v_{pec}(z)$ is given as a function of redshift $z$ as

$$v_{pec}(z) = c \frac{(1+z)^2 - 1}{(1+z)^2 + 1} \quad (1)$$

and this term may never exceed the speed of light [2]. It is only this velocity component that can have physical meaning and it is the rate of change of this velocity due to cosmological accelerations that could be responsible for the effect discussed here.

It is expected that, at any given moment, cosmological expansion is occurring uniformly throughout space on the largest scales. The expansion of the Universe is apparently accelerating in this current epoch, which is to say that clusters of galaxies separated by cosmological distances are receding from each other at an increasing rate. Since the large-scale cosmos is expanding spherically symmetrically with respect to any point in space, the distant masses are all currently accelerating radially outward from any point. However, even though the Universe should be expanding at the same accelerated rate everywhere, information about a body's acceleration should not be able to propagate through the Universe at a speed greater than the speed of light in a vacuum. A change in the acceleration of a cosmological mass located at a distance $R$ from a certain point should only be made evident to that point after a time $\sim R/c$. Thus the only possible interaction with the distant galaxies of the cosmos is with the galaxies as they were at an earlier time.

Detailed examination of the expansion history of the Universe has revealed that cosmological expansion has been accelerating since a look-back time corresponding to a redshift $z \sim 0.46$ [3]. In all prior cosmological eras the rate of cosmological expansion was decelerating. Therefore, galaxies whose redshifts are greater than $\sim 0.46$ with respect to some point are causally evident to that point as receding at a decelerating rate. As perceived from any location, most of the mass of the cosmos is situated at extremely remote distances with redshifts $z \gg 1$. The preponderance of cosmic mass is therefore causally manifest to any point in space as accelerating radially inward toward that point.

These cosmological accelerations should generate some inertial reaction in local masses. Only relative accelerations are meaningful, so the acceleration of the cosmic masses relative to local bodies should generate an inertial response in those local bodies. The nature of the putative inertial response to cosmological accelerations can be inferred from the properties that must be satisfied by any cosmological inertial induction. Regardless of its form, any interaction among masses that generates inertia must decrease as the distance between the masses increases. Also, the inertial reaction to a body's acceleration relative to the cosmic mass distribution must be generated in a direction opposite that of the relative acceleration [4].

Let there be a single point-like mass $m$, alone in space but for the very distant galaxies. Due to the large-scale homogeneity of the cosmos, the solitary mass could experience only a vanishing inertial reaction to the cosmological accelerations since there would be no preferred direction in which any reaction could occur. Now let there be a *pair* of non-interacting point-like masses $m$ separated by a distance $D$ much smaller than the radius of curvature of the Universe, alone but for the very distant cosmic masses. Let the azimuth of a cylindrical coordinate system pass through both test-masses and let the origin lie on the point exactly halfway between the two. In this scenario there would exist a preferred direction in which an inertial response to cosmological accelerations could be generated.

It is not readily apparent how to calculate in detail any supposed inertial reaction to retarded changes in the rate of cosmological expansion, particularly since the Universe is expanding spherically symmetrically relative to every point in space. However, due to the azimuthal symmetry of the hypothetical pair of test-masses $m$, it must be that the net effect of cosmological accelerations on the pair must be equivalent to the effect generated in a properly chosen one-dimensional analog. By symmetry, the mass distribution of the cosmos must behave identically to two point-like cosmological masses $M$ positioned on the azimuth on opposite sides of and a distance $R$ from the origin, where $R$ is of order the radius of curvature of the Universe (see Fig. 1). In order to represent the apparent decelerations of the preponderance of cosmic mass each mass *M* should be accelerating with some magnitude *a* toward the origin. The inertial response of each mass *m* to the accelerating cosmic masses *M* should be influenced more strongly by the cosmic mass to which it is closer. Each cosmic mass *M* would generate a relative acceleration between it and the nearest test-mass *m* such that the two are being accelerated toward each other. In other words, relative to each cosmological mass *M*, the nearest test-mass *m* is being accelerated outward, away from the origin and toward the mass *M*. Each test-mass *m* should therefore respond to the cosmological accelerations by experiencing an inertial induction in the direction away from the nearest cosmic mass *M*, toward the origin. This is to say that the test-masses *m* would effectively be attracted to each other in an inertial response to the cosmological accelerations.

This analysis can be extended to an ensemble of many particles. Consider some localized distribution of point-like masses alone in the cosmos but for the extremely remote galaxies and quasars. This could represent, for instance, a lone cluster of galaxies. The cosmological accelerations of the preponderance of mass in the Universe should induce an effective attraction between each pair of point-masses in the ensemble. The total force on each mass would depend on the ensemble's geometry, but in general the net response to cosmological accelerations should be a mutual attraction among the elements. Given the current inability to explain the accelerations observed in galaxies and clusters without introducing invisible matter or modifications of gravitational formulations, there is good motivation to investigate a possible unaccounted effect that generates attractive forces in ensembles of point-like masses.

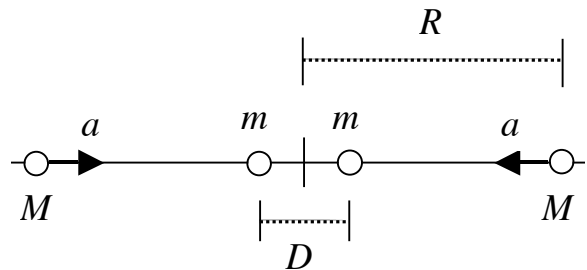

Fig. 1